\documentclass[preprint]{aastex}

\slugcomment{To appear in the {\em Astronomical Journal}}

\shorttitle{Chandra Observations of NGC~2139}
\shortauthors{Shields et al.}

\begin{document}


\title{Chandra Observations of the Nuclear Star Cluster and  \\
    Ultraluminous X-ray Sources in NGC~2139$^\ast$
\let\thefootnote\relax\footnotetext{$^\ast$Based in part 
on observations made with the NASA/ESA Hubble Space Telescope, and
obtained from the Hubble Legacy Archive, which is a collaboration
between the Space Telescope Science Institute (STScI/NASA), the
Space Telescope European Coordinating Facility (ST-ECF/ESA), and the
Canadian Astronomy Data Centre (CADC/NRC/CSA).}}


\author{Joseph C. Shields\altaffilmark{1}, Torsten B{\"o}ker\altaffilmark{2},
Luis C. Ho\altaffilmark{3}, Hans-Walter Rix\altaffilmark{4}, 
Roeland P. van der Marel\altaffilmark{5}, and C. Jakob Walcher\altaffilmark{6}}


\altaffiltext{1}{Department of Physics \& Astronomy, Ohio University, 
Clippinger Labs 251, Athens, OH 45701}
\altaffiltext{2}{European Space Agency, Keplerlaan 1, 200AG Noordwijk,
The Netherlands}
\altaffiltext{3}{The Observatories of the Carnegie Institution for
Science, 813 Santa Barbara St., Pasadena, CA 91101}
\altaffiltext{4}{Max-Planck-Institut f{\"u}r Astronomie, Konigstuhl
17, Heidelberg, 69117 Germany}
\altaffiltext{5}{Space Telescope Science Institute, 3700 San Martin
Dr., Baltimore, MD 21218}
\altaffiltext{6}{Leibniz-Institut f{\"u}r Astrophysik Potsdam (AIP), An der
Sternwarte 16, 14482 Potsdam, Germany}


\begin{abstract}

We report {\em Chandra} observations of the Scd galaxy NGC~2139,
which is known to host a recently formed ($10^{7.6}$ yrs) nuclear star
cluster.  The star cluster is undetected in X-rays, with an upper
bound on 0.5-7 keV luminosity of $L_{\rm X} < 7.1\times 10^{37}$ erg
s$^{-1}$.  This bound implies a bolometric accretion luminosity $<0.3$\%
of the Eddington luminosity for a black hole with the mass ($\sim
3400$ M$_\odot$) expected from extrapolation of the $M-\sigma$
relation.  The lack of X-ray emission indicates that a black hole, if
present, is not undergoing significant accretion at the current time.
While the central cluster is undetected, the data reveal a substantial
population of bright X-ray point sources elsewhere in this galaxy,
with eight qualifying as ultraluminous X-ray sources with $L_{\rm X} >
10^{39}$ erg s$^{-1}$.  We use archival {\em Hubble Space Telescope}
images to identify candidate optical counterparts for seven {\em
Chandra} sources, which in most cases have optical luminosities and
spatial profiles consistent with star clusters.  Compared with other
galaxies, the number of luminous X-ray sources in NGC~2139 is larger
by a factor of $\sim 4 - 10$ than expected based on its present star
formation rate and stellar mass.  This finding can be understood if
NGC~2139 has concluded a burst of star formation in the recent past,
and suggests that this galaxy could be important for testing the use
of X-ray source populations as a chronometer of star formation
history.

\end{abstract}


\keywords{galaxies: nuclei -- galaxies: star clusters: general -- 
galaxies:individual (NGC~2139) -- X-rays: galaxies -- X-rays: binaries}



\section{Introduction}

Surveys with the {\em Hubble Space Telescope} ({\em HST}\/) and other facilities
have demonstrated that the centers of galaxies commonly host a compact
star cluster, a supermassive black hole, or both.  Relative to star
clusters found elsewhere, nuclear clusters (NCs) are very dense, massive,
and often sites of multiple generations of star formation spanning a wide
range of ages \citep{walch05, walch06}.  NCs are found in galaxies of
all Hubble types.  The
sequence of events that lead to the formation of a NC is not well
established; the processes that drive the genesis and early growth of
supermassive black holes are similarly uncertain.  The location and
unusual properties of NCs invite speculation as to whether they have
a role in the formation and growth of a central black hole.  The
coexistence of NCs and black holes in some objects has been 
discussed by e.g., \citet{seth08} and \citet{grah09}.

An object of particular interest in this context is the Scd galaxy 
NGC~2139.  A NC in this system was discovered in the {\em HST}
snapshot survey of late-type galaxies by \citet{boek02}.  Follow-up
spectroscopic analysis was carried out by \citet{walch05, walch06} and
\citet{rossa06} who obtained a mass estimate for the NC of $8.3
\times 10^5$ M$_\odot$ and a stellar population dominated by stars
$\sim 4 \times 10^7$ years old.  Integral-field spectroscopy of the
inner part of NGC~2139 reported by \citet{ande08} indicated a
320-pc offset of the NC from the kinematic center of the galaxy.  The
NC's youth and location just outside the nucleus led \citeauthor{ande08}
to suggest that this system represents a NC in the making, as a
relatively young cluster formed in the inner disk settles into the
nucleus of this galaxy.

If the interpretation of NGC~2139 as a young NC system is correct, a
natural follow-on question is whether the NC hosts a significant black
hole.  The measurements by \citet{walch05, walch06} imply that the mass of any
black hole must be significantly less than the cluster stellar mass;
an analysis using Jeans models and the best fit stellar population
implies an upper limit of $M_\bullet <  1.5 \times 10^5$ M$_\odot$
\citep{neu12}.
An intermediate-mass black hole thus remains a possibility.  Such a
black hole could reveal itself through radiative signatures if it is 
undergoing accretion; consequently it is of interest to look for indicators
of an active galactic nucleus (AGN) associated with this system.
Evidence of weak AGN activity has been found in a growing number of
similar late-type galaxies with older NCs \citep[e.g.,][]{fil89, shie08,
satya09, desroc09, mcalp11}.

The optical spectra of the NC reported by \citeauthor{walch05} and
\citeauthor{ande08} show emission lines but do not require an accretion
source to explain them.  The emission-line luminosity is consistent
with the ionizing photon production rate predicted from the NC stellar
population analysis, and the line ratios are consistent with emission
from \ion{H}{2} regions.  The emission lines show rather broad wings
on some of the forbidden lines (full width at zero intensity $\sim
300$ km s$^{-1}$), but it seems likely that these are the result of
outflows driven by massive stars. If an AGN is present, it is outshone
in the optical bandpass by emission of stellar origin from the NC.

Measurements in the X-ray bandpass provide a sensitive alternative
means for detecting weak AGN activity, and for this reason we carried
out an observation of NGC~2139 using the {\em Chandra X-ray
Observatory}, which we report here.  The results have bearing on the
possible presence of a massive black hole in NGC~2139 and provide additional
information on energetic phenomena associated with this galaxy.  In
particular, the results reported here are of interest in 
providing constraints on the recent star formation history of this
galaxy, as traced by X-ray point source populations.  The number and
luminosities of the stellar sources provide strong constraints on the
amplitude and temporal progression of star formation activity.  
Extreme objects -- ultraluminous X-ray sources (ULXs) -- when present
are of unique interest in tracing the highest mass accretion sources
or other unusual phenomena emerging from a young stellar population.

\section{Observations}


We observed NGC~2139 with {\em Chandra} \dataset[ADS/Sa.CXO#obs/08196]
{(ObsID 8196)} using the ACIS instrument on 2007 June 17/18.  The
source was observed with the S3 chip with the detector operated in
faint mode, for a total exposure time of 38549 s.  The standard
pipeline was used for data processing and event screening.  Since we
are primarily interested in point sources in the inner portion of this
galaxy, we make use of measurements from this data set reported in the
{\em Chandra} \dataset[ADS/Sa.CXO#CSC] {Source Catalog release 1.1}
\citep[CSC;][]{evans10}.  Detection of CSC sources is based on the
WAVDETECT algorithm \citep{free02} with local background
determination.  The limiting detection significance employed
corresponds to $\sim 1$ false source over the central $\sim 16\farcm 9
\times 16\farcm 9$ field.

Within a 1\arcmin\ radius centered on the NGC~2139 nucleus, the CSC
lists 16 detected sources, with the properties presented in Table 1.
This search radius encompasses the bulk of the optical galaxy, which has
a disk scale length of $18\arcsec \approx 2.2$ kpc \citep{wein09}.  We
note that NGC 2139 is not a member of an identifiable galaxy group or cluster.
All of the detections are consistent with spatially unresolved point
sources; within the portion of the detector used, the point-spread
function has a 50\% enclosed energy fraction radius of $\sim 0\farcs
3$ over the energies observed.
The CSC provides fluxes for all of the sources assuming a
power-law spectrum with photon index $\Gamma = 1.7$, corrected for
absorption by the Galactic foreground HI column density of $4.1\times
10^{20}$ cm$^{-2}$ \citep{dickey90}.  Sources with at least
150 counts are also fit with a power-law spectrum with $\Gamma$ and
absorbing column density $N_{\rm H}$ left as free parameters.  Intrinsic
quantities listed in Table 1 are calculated assuming a distance of
23.6 Mpc as adopted in our earlier papers, which is based on a
Virgo-infall-corrected recession velocity with $H_0 = 70$ km s$^{-1}$
\citep{boek02}.  This distance falls at the low end of
redshift-independent distances listed on the NASA Extragalactic
Database, which span 24.0 to 36.3 Mpc for the same value of $H_0$.

\section{Results}

\subsection{The Nuclear Cluster}

The relationship between optical and X-ray sources in NGC~2139 can be
examined using archival {\em HST} images.  Two WFPC2 images exist
for this galaxy: the 
\dataset[ADS/Sa.HST#hst_08599_26_wfpc2_f814w_wf]{F814W image}
reported by \citet{boek02}, and a
\dataset[hst_08632_28_wfpc2_f300w_wf]{second image obtained with 
the F300W filter}.  Both images are available as part of the {\em
Hubble} Legacy Archive\footnote{http://hla.stsci.edu} (HLA), with
reprocessing that includes improved astrometry.  Astrometric solutions
for both the F814W and F300W images are available from 2MASS and GSC2
comparison sources in the field; astrometric solutions are computed
separately for each bandpass and for each set of comparison sources.
The HLA adopts the solution that minimizes residuals for the
comparison source positions, with the result that the F814W solution
is based on GSC2 sources and the F300W solution is based on the 2MASS
sources.  For both images, however, the GSC2 and 2MASS solutions are
systematically offset by similar amounts, with the result that HLA
astrometry for the F814W and F300W images is likewise offset;
cross-correlation of the two implies Right Ascension and Declination
for sources in the F300W image are larger by $\sim$0\farcs 31 and
$\sim$0\farcs 07, respectively, than in F814W. Given this ambiguity we
adopt the F814W solution along with its rms positional uncertainty of
0\farcs 4 in both RA and Dec, which is sufficient to encompass the
measured offset.

The X-ray source positions overlaid on the F814W image are shown in
Figure 1.  Based on the HLA {\tt daophot} source list for the F814W image,
the NC is located at RA$=06^h01^m07\fs88$, Dec$=-23\arcdeg 40\arcmin
20\farcs8$ (J2000.0), which differs in declination by +0\farcs 9 from
the position listed by \citet{ande08} based on the original
uncorrected astrometry.  {\em Chandra} source X9 is located near the NC and
might be considered a candidate counterpart.  The 95\% confidence
error radius for X9 is 0\farcs 45; assuming a Gaussian probability
profile, the 1-$\sigma$ uncertainty radius would be 0\farcs 18.
The NC and X9 are thus offset by $1\farcs 2 \pm 0\farcs 4$, which
suggests that the positions are genuinely different and the X-ray
source is {\em not} associated with the NC.  We note that adoption of
the F300W HLA astrometric solution would {\em increase} the separation
between the optical NC and X9.  Based on the coordinates provided by 
\citet{ande08}, shifted as noted above, source X9 is offset 
$3\farcs 9 \pm 0\farcs 3$ from the galaxy's kinematic center and
$2\farcs 1 \pm 1\farcs 3$ from the galaxy's photometric center. 

The {\em Chandra} data can be used to place an upper bound on the NC
X-ray luminosity.  Measurement of counts within an appropriate
aperture centered on the NC position is complicated by contamination
by X9.  We therefore measured counts within a 1\arcsec-radius
aperture, excluding the quadrant between position angles
$210-300^\circ$ where contamination is evident, and scaling by 4/3 to
correct for this missing area.  We measured the local background
between radii of 1\farcs 2 and 5\arcsec\ with regions selected to
avoid contamination from X9 and X10.  The result for the $0.5-7$ keV
bandpass is 4 counts in the NC aperture and average background for the
same area of 2.6 counts, which yields a Bayesian upper bound at 95\%
confidence of 6.8 counts for an actual source, assuming a constant
nonnegative prior and Poisson likelihood \citep{kraft91}.  We
translated this result into a flux using the exposure time and the
Portable Interactive Multi-Mission Simulator (PIMMS) version 3.9k.
Assuming absorption by the Galactic foreground HI column density of
$4.1\times 10^{20}$ cm$^{-2}$ \citep{dickey90} and a power-law
spectrum for the source with $\Gamma=2$, the result is $1.1 \times
10^{-15}$ erg s$^{-1}$ cm$^{-2}$, corresponding to an upper bound on
X-ray luminosity of $7.1\times 10^{37}$ erg s$^{-1}$.  If the NC
follows the correlation between stellar velocity dispersion
$\sigma_\ast$ and black hole mass $M_\bullet$ seen in other bulge-like
systems \citep{gebh00, ferrar00}, the measured $\sigma_\ast = 16.5\pm
2.5$ km s$^{-1}$ \citep{walch05} would predict $M_\bullet \approx
3400$ M$_\odot$ \citep{guelt09}\footnote{We note that the relationship
between $\sigma_\ast$ and $M_\bullet$ in this regime is highly
uncertain.}.  If we adopt a bolometric correction
appropriate for low-luminosity AGNs \citep{ho08} such that the
bolometric luminosity $L_{\rm bol} \approx 16 L_{\rm X}$, our upper bound
translates into $L_{\rm bol} < 1.1\times 10^{39}$ erg s$^{-1}$, or
$L_{\rm bol}/L_{\rm Edd} < 0.003$ where $L_{\rm Edd}$ is the Eddington luminosity.
A massive black hole, if present in the NC, does not manifest itself as a
significant X-ray source and is evidently undergoing little accretion.

An additional question of interest is whether our upper bound on the
luminosity of the NC is consistent with expectations based on its
stellar content.  As noted in \S 1, the NC is dominated by a young
population and thus can be expected to produce stellar remnants
generating significant X-ray emission.  Population models by
\citet{sipior03} for continuous star formation of 20 Myr duration
predict a $2-10$ keV luminosity of $\sim 2 \times 10^{37}$ erg
s$^{-1}$ for an object with the age and mass of the NC, corresponding to a
$0.5 - 7$ keV luminosity with $\Gamma = 2$ of $\sim 3 \times 10^{37}$
erg s$^{-1}$, consistent with the lack of detection and upper bound
reported above.

\subsection{X-ray Source Properties}

The X-ray sources that are detected in the Chandra observation are
noteworthy for their number and properties.  From Table 1, the
faintest detected X-ray source has a $0.5-7$ keV flux of $\sim 3\times
10^{-15}$ erg s$^{-1}$ cm$^{-2}$, which corresponds to $0.5-2$ keV and
$2-10$ keV fluxes of $1.3\times 10^{-15}$ and $2.3\times 10^{-15}$ erg
s$^{-1}$ cm$^{-2}$ respectively for a power-law spectrum with
$\Gamma=1.7$ typical of a quasar.  At these flux levels, X-ray surveys
predict an average of 0.6 and 1.0 sources within a circle of radius
1\arcmin\ on the sky \citep[e.g.,][]{capp07}, suggesting that
the large majority of the 16 sources found within the same area
centered on NGC~2139 are genuinely associated with that galaxy.

Luminosities calculated for the X-ray sources are listed in Table 1
and are in many cases large relative to typical X-ray binaries.
Adopting the usual threshold for ultraluminous X-ray (ULX) sources of
$10^{39}$ erg s$^{-1}$ implies that eight ULXs are present in NGC
2139.  ULXs are relatively rare and the number we find in NGC~2139 is
noteworthy; for comparison, the ULX catalog compiled by \citet{liu05}
lists only seven galaxies with eight or more ULXs.  ULX sources
have attracted attention since in at least some cases they seem to
require accretion that is anisotropic, in excess of the Eddington
limit, or powered by an intermediate-mass black hole.

Some insight into the spectral properties of the NGC~2139 sources can
be obtained from hardness ratios, defined here as $(M-S)/(M+S)$ and
$(H-M)/(H+M)$, where $S, M, H$ are source counts in bandpasses
following the CSC definitions of $0.5-1.2$, $1.2-2.0$, and $2.0-7.0$
keV, respectively.  Hardness ratios and uncertainties were derived
using the Bayesian procedure described by \citet{park06} and are
listed in Table 1.  The results are plotted in Figure 2 along with
predictions for power-law spectra generated with PIMMS. The
distribution of hardness ratios is similar to that found from other
studies of luminous X-ray binaries \citep[e.g.,][]{colb04, swartz04},
and in accord in most cases with power-law spectra
described by photon indices of $\Gamma \sim 2$.  The majority of
sources are consistent with significant absorption by column densities
$N > 10^{21}$ cm$^{-2}$, well in excess of the Galactic column, as
found also in previous ULX studies.  The impact of these parameter
values on measured flux is modest, as can be seen from comparison
of columns 8 and 11 in Table 1, listing flux estimates derived from
the default CSC spectral parameters and from fits with unconstrained
$\Gamma$ and $N_{\rm H}$.  While some investigations have suggested a link
between $\Gamma$ and $L_{\rm X}$ \citep[e.g.,][]{winter06}, no trend
between these quantities (indicated by the plotted curves and point
sizes, respectively) is evident in Figure 2.

The CSC applies several tests for variability to detected sources.  By
these criteria none of the NGC~2139 sources show evidence of
variability, with the exception of Sources 8 and 16 for the soft
($0.5-1.2$ keV) bandpass only.  Surveys of X-ray point sources of
comparable luminosity in other galaxies show similarly modest
evidence of variability over the timescales we sample \citep[e.g.,][]
{zezas06, grimm07, ghosh09}.  In the absence of variability the possibility exists
that a source listed in Table 1 could be a compact group of
multiple lower luminosity binaries that remains spatially unresolved
in the {\em Chandra} image.

\subsection{Optical Counterparts}

The {\em HST} images described in \S3.1 can be used to identify
candidate optical counterparts for the detected X-ray sources, and we
used the {\tt daophot} source lists from the HLA for this purpose.
{\tt Daophot}
source identification is optimized for point sources, with the result
that some highly extended structures are not identified (e.g., see X5
in Fig. 1).  Table 2
lists the results of source matching based on positional separations
$\leq 2\sigma$ (i.e. 95\% confidence) with errors in the optical and 
X-ray positions summed in quadrature.  Sources in the two WFPC2
bandpasses are matched if their positions agree within $<
0\farcs 05$, in which case a single entry with coordinates taken from
the F814W image is listed.  Three of the X-ray sources fall outside of
the field of view for the WFPC2 images, and 7 of the remaining 13 have
candidate optical counterparts.  Five of the X-ray sources have more
than one optical match; sources X8, X9, X10, and X11 fall along the
central optical bar of the galaxy, where multiple clusters and knots
of optical emission that likely trace recent star formation are
evident.  X12 is in a similarly complex region, resulting in six
possible associations among the optical sources.  Most of the
candidate counterparts are detected in only one of the optical
bandpasses.

The potential counterparts show diverse photometric properties,
consistent with results from earlier studies of luminous X-ray sources
in galaxies \citep[e.g.,][]{ptak06, ghosh09}.  Photometry listed in
Table 2 is given in AB magnitudes measured in apertures with radius
0\farcs 15.  For comparison, the equivalent {\tt daophot} measurements for
the NC are $18.79\pm 0.01$ and $18.31\pm 0.01$ in F300W and F814W,
respectively.  The sources listed in Table 2 span a considerable range
of brightness, and those detected in both bandpasses show a
substantial range in color.  Given NGC~2139's distance modulus of 31.9
mag, most of the optical sources are probably too luminous to be
single stars; the brightest known stellar counterparts to ULXs in other
galaxies have absolute magnitudes of $M_V \approx -7$ to $-8$
\citep{robe08,gris12} corresponding in the present case to apparent magnitudes of $m_V
\approx m_{AB} \approx 24 - 25$, well below our detection threshold.
The optical sources in Table 2 are more likely star clusters.  Supporting evidence for
this conclusion is provided by the HLA {\tt daophot} measurements, which
flag the majority of the candidate counterparts as extended (Table 2)
based on a concentration index.  We note that as is often the case in
studies of this type, there is considerable possibility for confusion
within the existing data and it is difficult to assign a definitive
correspondence between the X-ray sources and any of the objects listed
in Table 2.  Aside from their number, the ULXs in NGC~2139 have
observational properties to the extent we can measure them
that are similar to those for sources found elsewhere, and the same
physical scenarios for their interpretation apply (\S 3.2).

\subsection{The Host Galaxy and the X-ray Source Population}

A number of studies have reported evidence for a link between star
formation and the incidence of luminous X-ray point sources in
galaxies \citep[e.g.,][]{erac02, kilg02, swartz04, gilf04, liu06},
which can be understood as the
manifestation of X-ray binaries created when massive stars in binary
systems evolve into compact remnants that undergo accretion.  
\citet{grimm03} have argued that the resulting luminosity function of X-ray
point sources can be used as an indicator of star formation rate in
galaxies undergoing significant star formation.  The cumulative
luminosity function for the sources listed in Table 1 is shown in
Figure 3, along with the universal luminosity function from
\citeauthor{grimm03} depicted with several scalings, suggesting 
a star formation rate of $\sim 5 - 7$ M$_\odot$ yr$^{-1}$.

Independent estimates of the star formation rate in NGC~2139 are
available from measurements in other bandpasses.  We calculated the
far-infrared (FIR) luminosity $L_{\rm FIR}$ using {\em IRAS} 60 and 100
$\mu$m fluxes \citep{sand03} and the formula given by \citet{rice88},
yielding $L_{\rm FIR} = (2.70 \pm 0.01)\times 10^{43}$ erg s$^{-1}$.  We also
obtained the H$\alpha$ luminosity using the integrated flux, corrected
for foreground Galactic absorption, measured by \citet{moust06}, with
the result that $L_{H\alpha} = (2.20 \pm 0.09) \times 10^{41}$ erg s$^{-1}$.
Based on the proportional relationships reported by \citet{kenn98},
$L_{\rm FIR}$ and $L_{H\alpha}$ imply star formation rates of 1.2 and 1.7
M$_\odot$ yr$^{-1}$.  Uncertainties in these estimates are likely dominated
by systematic errors, and \citet{kenn98} notes that different calibrations
of the scaling relationships vary by $\sim 30$\%.  Allowing for such
uncertainties, both rates are lower than suggested by the X-ray source
counts using the \citet{grimm03} relation.

The star formation rates
obtained from $L_{\rm FIR}$ and $L_{H\alpha}$ may be too low if significant
reprocessed emission emerges outside the FIR bandpass, or if H$\alpha$
is significantly attenuated by dust.  Both of these considerations can
be addressed using a formulation provided by \citet{kenn09} for
correcting H$\alpha$ luminosities for internal extinction using the
total infrared (TIR) luminosity as an indicator of reprocessing by
dust.  We used their equation (12) with the TIR luminosity derived from
{\em IRAS} 25, 60, and 100 $\mu$m fluxes using equation (5) from
\citet{dale02} to obtain a corrected
$L_{H\alpha} = (3.6 \pm 0.3) \times 10^{41}$ erg s$^{-1}$, implying a star
formation rate of $2.8 \pm 0.2$ M$_\odot$ yr$^{-1}$, or $2.8 \pm
0.8$ M$_\odot$ yr$^{-1}$ if a 30\% calibration error is adopted.
This result is still rather low compared with the \citet{grimm03}
findings.  We note that all of these
formulations for estimation of star formation rate assume a standard
stellar initial mass function.

An older stellar population may contribute an important fraction of
the observed X-ray sources.  \citet{colb04} obtained
relationships predicting the total point-source X-ray luminosity
$L_{\rm XP}$ in merger and spiral galaxies with an old-population
contribution scaled from the galaxy's $K$-band luminosity $L_K$.  
The NGC~2139 measurement by 2MASS \citep[20 mag arcsec$^{-2}$ 
isophotal magnitude]{skrut06} implies $L_K =
8.88 \times 10^{42}$ erg s$^{-1}$,  which we combined with $L_{\rm FIR}$
and \citeauthor{colb04}'s equation (4) to predict $L_{\rm XP} = 2.1 \times
10^{39}$ erg s$^{-1}$ ($0.5 - 8$ keV).  Summing the luminosities for
the point sources listed in Table 1 yields a measured $L_{\rm XP} = 2.1
\times 10^{40}$ erg s$^{-1}$, a factor of 10 larger than the predicted
value.  In estimating the young stellar population component, 
\citeauthor{colb04} use an $L_{\rm FIR}$ value that is increased to include an estimate
of unabsorbed ultraviolet luminosity derived from fitting the galaxy's
spectral energy distribution.  We did not attempt this correction, but
note that for the galaxies tabulated by \citeauthor{colb04} the UV contribution
increases the luminosity of the young component by at most a factor of
2, which in the case of NGC~2139 would predict $L_{\rm XP}$ of at most
$3.5 \times 10^{39}$ erg s$^{-1}$, still a factor of 6 below the
observed value.  We note that uncertainties in the distance to
NGC~2139 do not affect these arguments since they rely on ratios of
luminosity in different bandpasses; since our adopted distance is
conservatively low (\S 2), we have if anything underestimated the
number of ULXs.  As a point of reference, we can follow \citet{colb04} in
estimating the stellar mass from the galaxy's $B$ magnitude of
12.0 \citep{deva91}, the resulting color $B-K = 2.7$, and mass-to-light
ratio of $\log (M/L)_K = -0.387$ \citep{bell01}, yielding $M = 7.1
\times 10^9$ M$_\odot$. To summarize the foregoing discussion, the observed
number and luminosity of X-ray point sources in NGC~2139 is high in
relation to other indicators of star formation or stellar mass, based
on the trends seen in other galaxies.  

The unusual number of luminous X-ray sources in NGC~2139 may imply
that this galaxy is moving through a distinct phase when a major
episode of star formation recently occurred but is in the process of
shutting off.  \citet{colb04} found a small set of starburst
galaxies with $L_{\rm XP}$ that exceeded their predicted values by up to a
factor of $\sim 4$.  With $L_{\rm FIR}/L_K = 3.0$, NGC~2139 has attributes
of a starburst, and the excess $L_{\rm XP}$ is thus not surprising.
Inspection of Digital Sky Survey images of this galaxy reveals an
extension to the southeast that is suggestive of a recent interaction
that might have triggered the starburst.  Population synthesis models
predict a maximum $L_{\rm XP}$ after the conclusion of a burst of star
formation \citep{vanbever00, sipior03}, and
spatial offsets between ULXs and \ion{H}{2} regions may similarly
trace a lag between peak star formation and maximum accretion
luminosity \citep{swartz09}.  The large number of luminous X-ray
sources in NGC~2139 thus suggests that this galaxy could be in the final
stages of a starburst.  Alternatively, given that a large fraction of
the detected point sources are ULXs and hence relatively exotic
sources, NGC 2139 may probe a regime where the scaling relationships
between star formation and normal high-mass X-ray binary
behavior break down.  In either circumstance, this galaxy presents a valuable
test case for use of the X-ray source population as a star formation
chronometer and for understanding the origins of ULXs.

\section{Conclusions}

The young NC in NGC~2139 does not show X-ray evidence for having
formed a significant black hole.  However, the NC may share a link to
the many luminous extranuclear X-ray sources in NGC~2139 in that they
are the products of a major starburst event within the past $\sim
10^8$ yr which is now subsiding.  This example raises the question of
whether an enhanced episode of large-scale star formation is important
more generally for generation of NCs and their unusual properties.  In
NGC~2139 the question remains open as to what triggered a starburst.
The large number of luminous X-ray sources relative to other
indicators of star formation suggests that this galaxy may have
unusual value for studying the temporal progression of radiative
starburst signatures.

\acknowledgments

We thank Mike Eracleous for informative discussions.  This research
has made use of data obtained from the Chandra Source Catalog,
provided by the Chandra X-ray Center (CXC) as part of the Chandra Data
Archive.  This study has additionally made use of the NASA/IPAC
Extragalactic Database (NED) which is operated by the Jet Propulsion
Laboratory, California Institute of Technology, under contract with
NASA.  Support for this work was provided by NASA through {\em
Chandra} Award Number GO7-8088X issued by the CXC, which is operated
by the Smithsonian Astrophysical Observatory for and on behalf of NASA
under contract NAS8-03060.

{\it Facilities:} \facility{CXO (ACIS)}, \facility{HST (WFPC2)}.




\clearpage

\begin{figure}
\plotone{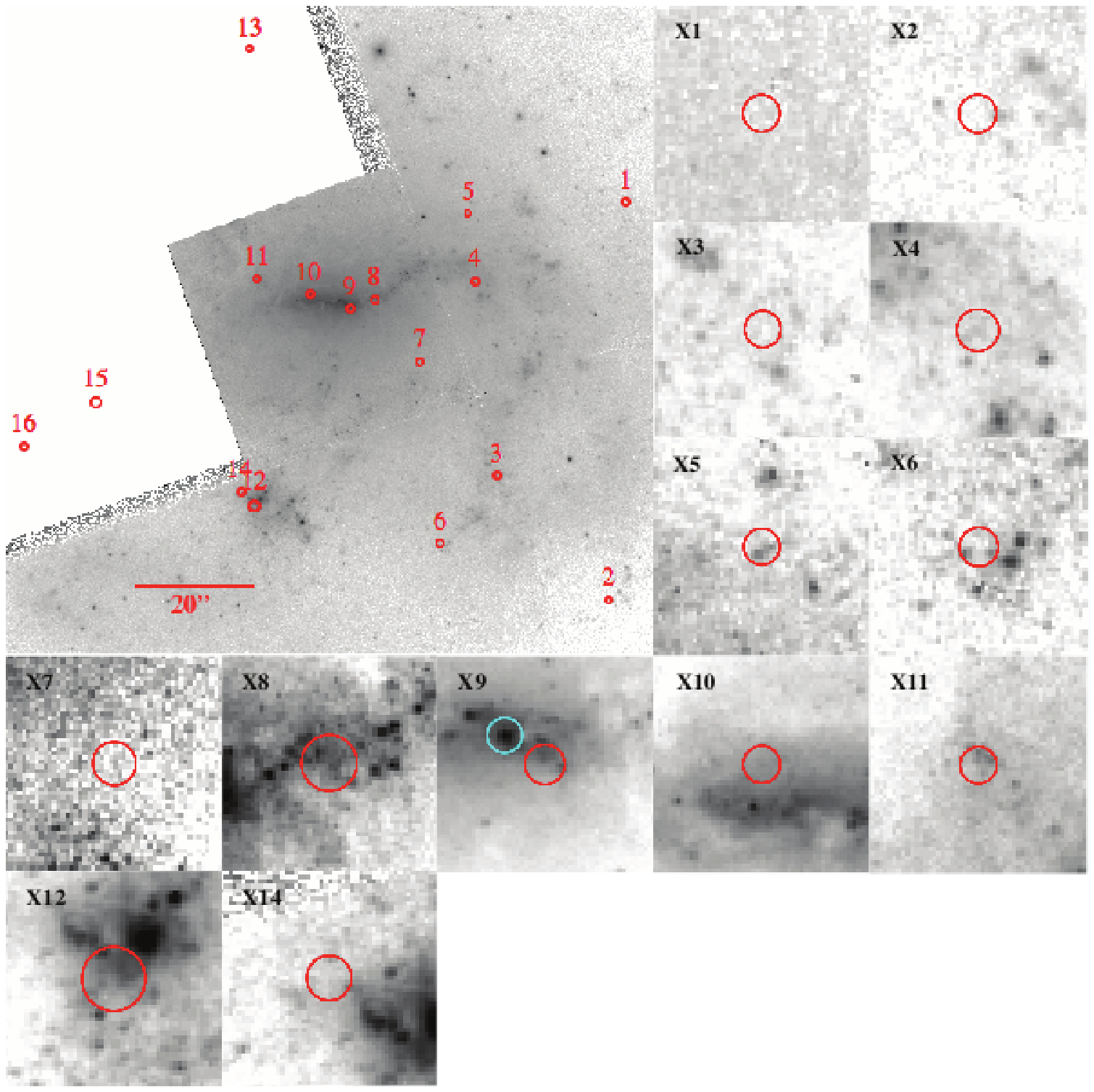}

\caption{HST WFPC2 F814W image of NGC~2139 with X-ray source positions
indicated (upper left); north is up and east to the left.  The
numbered boxes show zoomed regions (5\arcsec on a side) around each of the
X-ray source positions indicated by red circles, with radius equal to 
the astrometric  uncertainty from Table 1.  The cyan circle in the X9 box 
indicates the location of the NC.
\label{fig1}}
\end{figure}

\begin{figure}
\plotone{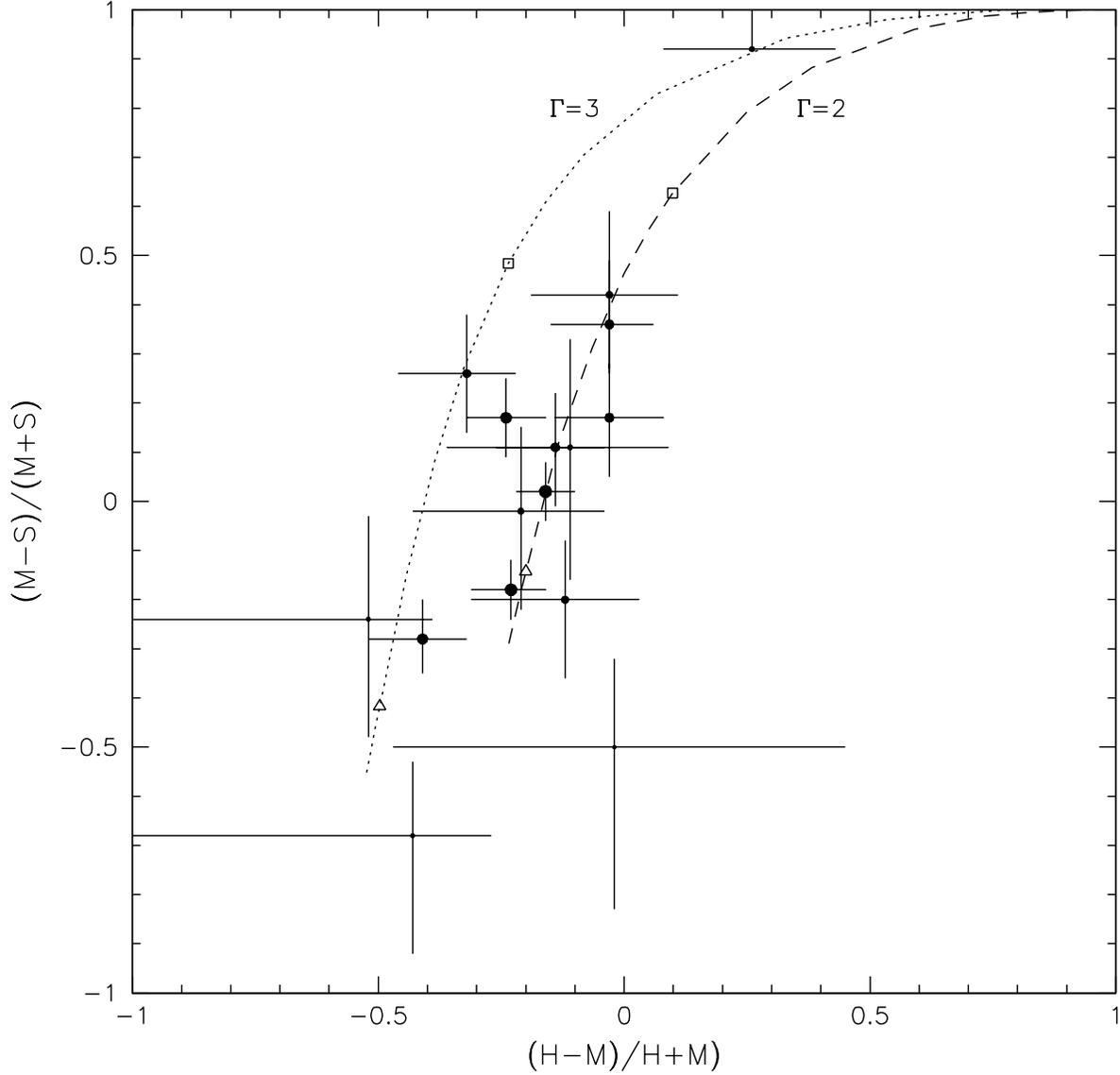}
\caption{Hardness ratios for the X-ray sources; point sizes scale
logarithmically with source luminosity.  The plotted curves show
predictions for power-law spectra, as a function of intervening
absorption column density ranging from $N_{\rm H} = 0$ to $10^{23}$
cm$^{-2}$.  The open triangles and squares indicate results for
$N_{\rm H} = 10^{21}$ and $10^{22}$ cm$^{-2}$, respectively.
\label{fig2}}
\end{figure}

\begin{figure}
\plotone{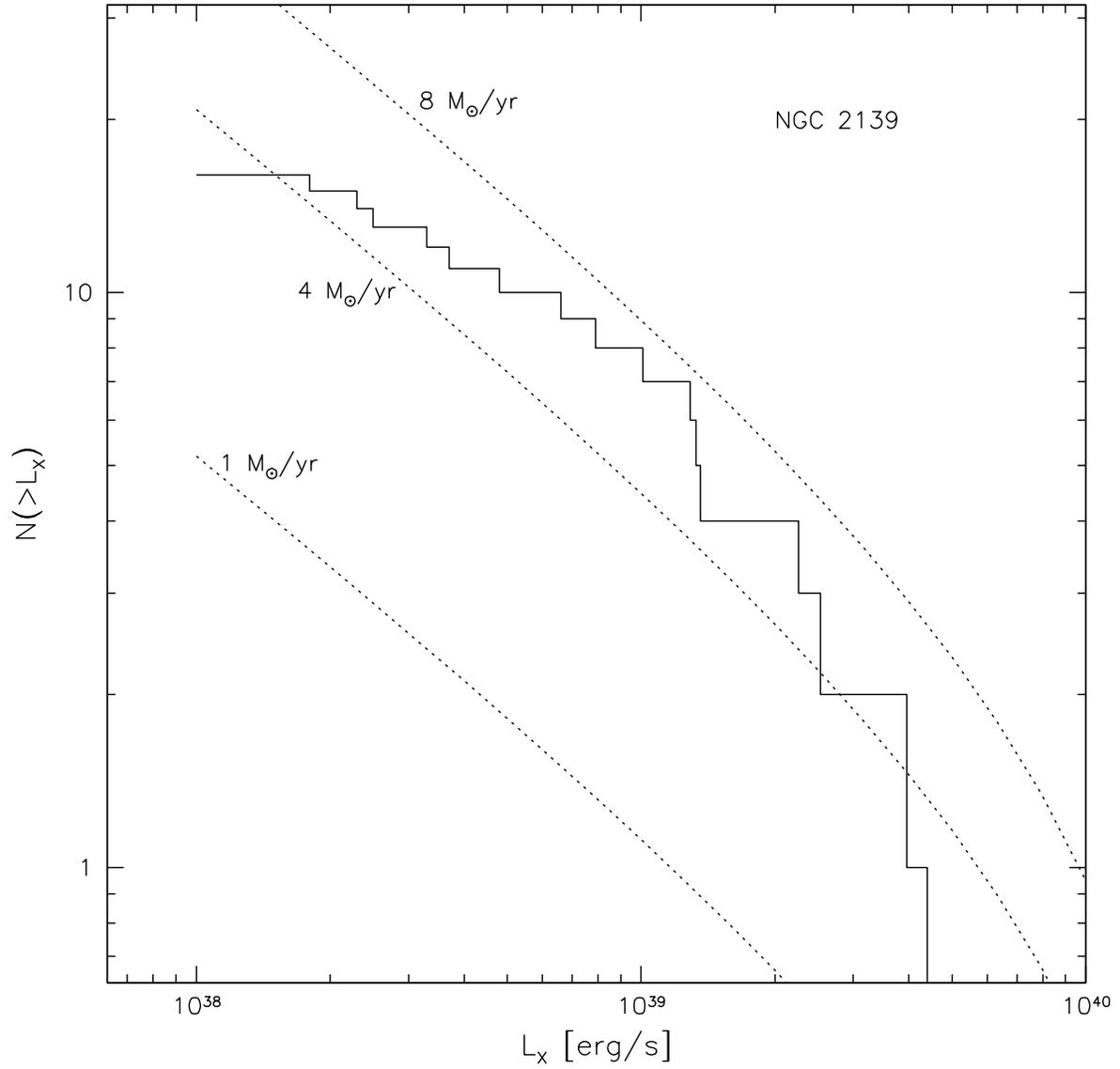}
\caption{Cumulative luminosity function for the X-ray sources
in NGC~2139.  The curves show predictions for the indicated
star formation rate from \citet{grimm03}.
\label{fig3}}
\end{figure}

\clearpage

\begin{deluxetable}{crrrrrrrrcrrr}
\tabletypesize{\scriptsize}
\rotate
\tablecaption{X-ray Source Properties\label{tbl-1}}
\tablewidth{0pt}
\tablehead{
\colhead{No.} & \colhead{R.A.} & \colhead{Dec.} & \colhead{$\sigma_r$} & 
\colhead{Counts} & \colhead{$(M-S)/(M+S)$} & \colhead{$(H-M)/(H+M)$} &
\colhead{$f_{PL}$} & \colhead{$L_{\rm X}$} & \colhead{$f_{\rm fit}$} & \colhead{$\Gamma$} &
\colhead{$N_{\rm H}$} & \colhead{$L_{\rm fit}$} \\
\colhead{(1)} & \colhead{(2)} & \colhead{(3)} & \colhead{(4)} & \colhead{(5)} & \colhead{(6)} & \colhead{(7)} & \colhead{(8)} & \colhead{(9)} & \colhead{(10)} & \colhead{(11)} & \colhead{(12)} & \colhead{(13)} 
}
\startdata
 1 & 6 01 05.02 & $-23$ 40 06.76 & 0.42 & 200 & $ -0.28_{-0.07}^{+0.08}$ & $ -0.41_{-0.11}^{+0.09}$  & $ 3.4  \pm 0.2   $ & $ 23  \pm 2   $ & $2.2^{+0.2}_{-0.1}$ & $2.5^{+0.6}_{-0.4}$     & $4^{+11}_{-4}$   & $14.9^{+1.7}_{-0.8}$ \\
 2 & 6 01 05.19 & $-23$ 41 01.99 & 0.43 & 115 & $  0.36_{-0.10}^{+0.13}$ & $ -0.03_{-0.12}^{+0.09}$  & $ 2.0  \pm 0.2   $ & $ 13  \pm 1   $ &                   &                        &                     & \\
 3 & 6 01 06.33 & $-23$ 40 44.78 & 0.42 & 119 & $  0.11_{-0.12}^{+0.11}$ & $ -0.14_{-0.12}^{+0.10}$  & $ 2.0  \pm 0.2   $ & $ 13  \pm 1   $ &                   &                        &                     & \\
 4 & 6 01 06.54 & $-23$ 40 17.63 & 0.48 & 42  & $ -0.02_{-0.20}^{+0.17}$ & $ -0.21_{-0.22}^{+0.17}$  & $ 0.72 \pm 0.12  $ & $ 4.8 \pm 0.8 $ &                   &                        &                     & \\
 5 & 6 01 06.61 & $-23$ 40 08.33 & 0.41 & 351 & $ -0.18_{-0.06}^{+0.06}$ & $ -0.23_{-0.08}^{+0.07}$  & $ 6.0  \pm 0.3   $ & $ 40  \pm 2   $ & $4.6^{+0.4}_{-0.3}$ & $2.1^{+0.3}_{-0.3}$      & $11^{+7}_{-7}$   & $30^{+2}_{-2}$ \\
 6 & 6 01 06.90 & $-23$ 40 54.13 & 0.44 & 89  & $  0.26_{-0.12}^{+0.12}$ & $ -0.32_{-0.14}^{+0.10}$  & $ 1.5  \pm 0.2   $ & $ 10  \pm 1   $ &                   &                        &                     & \\
 7 & 6 01 07.10 & $-23$ 40 28.88 & 0.49 & 29  & $  0.11_{-0.27}^{+0.22}$ & $ -0.11_{-0.25}^{+0.20}$  & $ 0.50 \pm 0.10  $ & $ 3.3 \pm 0.7 $ &                   &                        &                     & \\
 8 & 6 01 07.56 & $-23$ 40 20.30 & 0.64 & 16  & $ -0.50_{-0.33}^{+0.18}$ & $ -0.02_{-0.45}^{+0.47}$  & $ 0.27 \pm 0.08  $ & $ 1.8 \pm 0.5 $ &                   &                        &                     & \\
 9 & 6 01 07.81 & $-23$ 40 21.51 & 0.45 & 70  & $ -0.20_{-0.16}^{+0.12}$ & $ -0.12_{-0.19}^{+0.15}$  & $ 1.2  \pm 0.2   $ & $ 7.9 \pm 1.0 $ &                   &                        &                     & \\
10 & 6 01 08.22 & $-23$ 40 19.53 & 0.42 & 120 & $  0.17_{-0.12}^{+0.11}$ & $ -0.03_{-0.11}^{+0.11}$  & $ 2.0  \pm 0.2   $ & $ 14  \pm 1   $ &                   &                        &                     & \\
11 & 6 01 08.75 & $-23$ 40 17.44 & 0.41 & 390 & $  0.02_{-0.06}^{+0.06}$ & $ -0.16_{-0.06}^{+0.06}$  & $ 6.6  \pm 0.4   $ & $ 44  \pm 2   $ & $5.8^{+0.3}_{-0.3}$   & $2.0^{+0.2}_{-0.2}$       & $21^{+7}_{-7}$    & $38^{+2}_{-2}$ \\
12 & 6 01 08.78 & $-23$ 40 48.92 & 0.73 & 20  & $ -0.68_{-0.24}^{+0.15}$ & $ -0.43_{-0.57}^{+0.16}$  & $ 0.34 \pm 0.09  $ & $ 2.3 \pm 0.6 $ &                   &                        &                     & \\
13 & 6 01 08.83 & $-23$ 39 45.57 & 0.42 & 224 & $  0.17_{-0.08}^{+0.08}$ & $ -0.24_{-0.08}^{+0.08}$  & $ 3.8  \pm 0.3   $ & $ 25  \pm 2   $ & $3.5^{+0.4}_{-0.5}$ & $2.0^{+0.5}_{-0.4}$     & $32^{+16}_{-15}$  & $23^{+3}_{-3}$ \\
14 & 6 01 08.92 & $-23$ 40 46.96 & 0.51 & 33  & $  0.92_{-0.01}^{+0.08}$ & $  0.26_{-0.18}^{+0.17}$  & $ 0.56 \pm 0.10  $ & $ 3.7 \pm 0.7 $ &                   &                        &                     & \\
15 & 6 01 10.38 & $-23$ 40 34.59 & 0.58 & 22  & $ -0.24_{-0.24}^{+0.21}$ & $ -0.52_{-0.48}^{+0.13}$  & $ 0.37 \pm 0.08  $ & $ 2.5 \pm 0.6 $ &                   &                        &                     & \\
16 & 6 01 11.11 & $-23$ 40 40.63 & 0.48 & 58  & $  0.42_{-0.15}^{+0.17}$ & $ -0.03_{-0.16}^{+0.14}$  & $ 0.99 \pm 0.14  $ & $ 6.6 \pm 0.9 $ &                   &                        &                     & \\
\enddata

\tablecomments{
Col. (1): X-ray source number. Cols. (2) and (3): J2000 coordinates. 
Col. (4): 95\%-confidence positional uncertainty radius in arcseconds. 
Col. (5): X-ray counts, $0.5-7$ keV.  Cols. (6) and (7): Hardness ratios 
and 95\% confidence error bars. Col. (8): Flux in units of $10^{-14}$ 
erg s$^{-1}$ cm$^{-2}$ ($0.5-7$ keV), assuming a power-law spectrum with 
$\Gamma = 1.7$ corrected for foreground Galactic absorption.  
Col. (9): Luminosity ($0.5-7$ keV) in units of $10^{38}$ erg s$^{-1}$ derived
from flux in Col. (8).
Col. (10) -- (13): Results of power-law spectral fit with $\Gamma$ and
$N_{\rm H}$ (in units of $10^{20}$ cm$^{-2}$) as free parameters; flux and 
luminosity with units as in Cols. (8) and (9).
}
\end{deluxetable}

\begin{deluxetable}{crrrrrrrc}
\tabletypesize{\scriptsize}
\tablecaption{Optical Counterpart Candidates\label{tbl-2}}
\tablewidth{0pt}
\tablehead{
\colhead{No.} & \colhead{R.A.} & \colhead{Dec.} & \colhead{$\Delta r$} & 
\colhead{$m_{814}$} & \colhead{$\sigma_{814}$} & \colhead{$m_{300}$} & \colhead{$\sigma_{300}$} & \colhead{Extended} \\
\colhead{(1)} & \colhead{(2)} & \colhead{(3)} & \colhead{(4)} & \colhead{(5)} & \colhead{(6)} & \colhead{(7)} & \colhead{(8)} 
}
\startdata

 2 & 6 01 05.15 & $-23$ 41 02.05 & 0.54 & 24.05 & 0.08 &       &      & N \\
 3 & 6 01 06.28 & $-23$ 40 44.40 & 0.78 &       &      & 23.12 & 0.06 & Y \\
 6 & 6 01 06.85 & $-23$ 40 54.45 & 0.78 & 22.52 & 0.03 & 22.06 & 0.02 & Y \\
   & 6 01 06.92 & $-23$ 40 54.29 & 0.36 & 23.92 & 0.09 &       &      & N \\
 8 & 6 01 07.51 & $-23$ 40 19.87 & 0.81 &       &      & 21.95 & 0.07 & Y \\
   & 6 01 07.60 & $-23$ 40 19.81 & 0.74 &       &      & 21.48 & 0.05 & Y \\
 9 & 6 01 07.78 & $-23$ 40 21.56 & 0.38 & 19.79 & 0.04 & 21.01 & 0.04 & Y \\
   & 6 01 07.81 & $-23$ 40 20.99 & 0.52 & 19.18 & 0.02 & 20.21 & 0.02 & Y \\
   & 6 01 07.80 & $-23$ 40 21.28 & 0.25 &       &      & 20.37 & 0.02 & Y \\
   & 6 01 07.82 & $-23$ 40 21.18 & 0.36 &       &      & 20.17 & 0.02 & Y \\
11 & 6 01 08.73 & $-23$ 40 17.26 & 0.31 & 21.50 & 0.05 &       &      & Y \\
   & 6 01 08.75 & $-23$ 40 17.17 & 0.27 & 21.48 & 0.05 &       &      & Y \\
   & 6 01 08.74 & $-23$ 40 17.62 & 0.21 &       &      & 22.14 & 0.06 & Y \\
12 & 6 01 08.80 & $-23$ 40 48.32 & 0.65 & 20.62 & 0.03 & 21.09 & 0.04 & Y \\
   & 6 01 08.76 & $-23$ 40 48.90 & 0.27 & 20.95 & 0.04 &       &      & Y \\
   & 6 01 08.76 & $-23$ 40 48.83 & 0.26 &       &      & 20.91 & 0.03 & Y \\
   & 6 01 08.80 & $-23$ 40 49.73 & 0.88 & 21.67 & 0.05 &       &      & N \\
   & 6 01 08.73 & $-23$ 40 49.26 & 0.80 &       &      & 21.77 & 0.05 & Y \\
   & 6 01 08.76 & $-23$ 40 48.59 & 0.38 &       &      & 20.64 & 0.03 & Y \\
\enddata
\tablecomments{
Col. (1): X-ray source number. Col. (2)-(3): Optical position, J2000.
Col. (4): Offset between X-ray and optical positions in arcseconds. 
Col. (5)-(8): Optical AB magnitudes and 1$\sigma$ uncertainties.
Col (9): Whether source is extended, based on difference in magnitudes for 
aperture radii of 0\farcs 05 and 0\farcs 15.
}
\end{deluxetable}

\clearpage

\end{document}